\documentclass[a4paper,11pt]{article}
\usepackage{pos}
\usepackage{natbib}
\let\OLDthebibliography\thebibliography
\renewcommand\thebibliography[1]{
  \OLDthebibliography{#1}
  \setlength{\parskip}{0pt}
  \setlength{\itemsep}{-1pt}
}

\title{Resolved photoproduction in \texttt{MadGraph5\_aMC@NLO}}

\author*[a]{Laboni Manna}    
\author[a]{Anton Safronov}
\author[b,c]{Carlo Flore}
\author[a]{Daniel Kikola}
\author[d]{Jean-Philippe Lansberg}
\author[e]{Olivier Mattelaer}

\affiliation[a]{Warsaw University of Technology, plac Politechniki 1, Warsaw, Poland}
\affiliation[b]{Dipartimento di Fisica, Università di Cagliari, Cittadella Universitaria, I-09042 Monserrato (CA), Italy}
\affiliation[c]{INFN, Sezione di Cagliari, Cittadella Universitaria, I-09042 Monserrato (CA), Italy}
\affiliation[d]{Université Paris-Saclay,CNRS, IJCLab, 91405 Orsay, France}
\affiliation[e]{Centre for Cosmology, Particle Physics and Phenomenology (CP3), 
Université Catholique de Louvain, Chemin du Cyclotron, Louvain-la-Neuve, B-1348, Belgium}

\emailAdd{laboni.manna.dokt@pw.edu.pl}
\emailAdd{anton.safronov.dokt@pw.edu.pl}
\emailAdd{carlo.flore@unica.it}
\emailAdd{daniel.kikola@pw.edu.pl}
\emailAdd{Jean-Philippe.Lansberg@in2p3.fr}
\emailAdd{ olivier.mattelaer@uclouvain.be}

\abstract{The upcoming Electron-Ion Collider (EIC), with its high luminosity, will offer an unprecedented opportunity to explore the internal structure of atomic nucleus over an extended energy range from  $\sqrt{s_{ep}} =$ 45 GeV to  $\sqrt{s_{ep}} =$ 140 GeV. A particularly promising aspect of this collider is the study of the partonic structure with quasi-real photons which can also be studied in inclusive ultra-peripheral collisions at the Large Hadron Collider (LHC). In this work, we present our validation of resolved photoproduction at fixed order for (next-to-)leading order using \texttt{MadGraph5\_aMC@NLO}, a widely adopted framework at the LHC.}

\FullConference{31st International Workshop on Deep Inelastic Scattering (DIS2024)\\
 8–12 April 2024\\
Grenoble, France\\}

\begin{document}
\maketitle

\section{Introduction}\vspace*{-0.3cm}
In lepton-proton collisions, the cross section mainly results from processes where the lepton emits a quasi-real photon, which then interacts with the constituents of the proton. Such processes are referred to as photoproduction. Two types of processes contribute to photoproduction. Direct-photon processes involve a photon engaging directly in a hard subprocess with a parton from the proton, whereas resolved-photon processes involve a photon acting as a source of partons, with one of these partons participating in the hard subprocess.

Resolved-photon processes are particularly interesting because the photon, due to quantum fluctuations, manifests a hadronic structure. This structure allows partons from both the photon and the proton to interact via the strong interaction. These events are crucial for studying the partonic structure of the photon. Its exploration has a long history, dating back to theoretical works in 1971~\cite{PhysRevLett.27.280}. The upcoming Electron-Ion Collider (EIC)~\cite{Boer:2024ylx,ABDULKHALEK2022122447,Accardi:2012qut}, which will provide high luminosities for polarised electron and proton collisions, is an outstanding tool for studying the partonic structure of the photon with high precision. Moreover, it offers the opportunity to explore the previously unknown polarised photon Parton Distribution Functions (PDFs).

While several tools are currently available for direct photoproduction, such as \texttt{HELAC-Onia}~\cite{Shao:2015vga}, \texttt{Pythia}~\cite{Helenius:2017aqz}, and single-use codes like \texttt{FMNR}~\cite{Frixione:1993dg}, they all have limitations. In this work, we will demonstrate our validation of resolved photoproduction at next-to-leading order (NLO) for fixed-order mode using \texttt{MadGraph5\_aMC@NLO} ({\tt MG5})~\cite{Alwall:2014hca} which is capable of automatically computing NLO results, as recently done for \texttt{SHERPA}~\cite{Hoeche:2023gme,Meinzinger:2023xuf}, for
the upcoming EIC, but also for inclusive ultra-peripheral proton-nucleus and nucleus-nucleus collisions at the LHC (see e.g.~\cite{Lansberg:2024zap}).

\section{Framework}\vspace*{-0.3cm}
According to the collinear QCD factorisation theorem, the cross section for the scattering of two hadrons, $h$, resulting in the production of a specific observed final state $C$ plus other unobserved particles $X$, can be expressed as a convolution of two components: a perturbatively calculable partonic cross section and non-perturbative PDFs of the hadrons. In \texttt{MG5}, it can be summarised as:
\begin{equation}
    \sigma_{hh\to C+X}=\sum\limits_{a,b} \int\!dx_{a} dx_{b}{f_{a}^{h} (x_{a},\mu_{F}; {\tt LHAID\_h}) }{f_{b}^{h} (x_{b},\mu_{F}; {\tt LHAID\_h}) }{\hat{{\sigma}}}_{ab \to C+X}(x_{a},x_{b},\mu_{F},\mu_{R})
    \label{eq:equation1}
\end{equation}
where $x_{a,b}$ represent the momentum fractions carried by the partons (either gluons or quarks) originating from the hadrons, $\mu_R$ and $\mu_F$ denote the renormalisation and factorisation scales respectively, $f_{a}^{h}$ and $f_{b}^{h}$ are PDFs of the incoming hadrons with $\texttt{LHAPDF}$ set id $\texttt{LHAID\_h}$, and $\hat{\sigma}_{ab \to C+X}$ is the partonic cross section for the process. Eq.~(\ref{eq:equation1}) serves as the fundamental equation for the development of \texttt{MG5} and is specific for symmetric \textit{hh} collisions with the same \texttt{LHAID\_h}. As an upgrade to \texttt{MG5}, we have extended its capabilities to include two different types of asymmetric collisions: asymmetric hadron-hadron collisions and photoproduction between a charged particle and a hadron. In the first scenario, by modifying the existing algorithm of \texttt{MG5}, we enable the simultaneous invocation of two distinct \texttt{LHAPDF} sets to compute the corresponding cross section~\cite{Safronov:2022gr} as: \vspace*{-0.2cm}
\begin{equation} 
    \sigma_{AB\to C+X}=\sum\limits_{a,b} \int\!dx_{a} dx_{b}{f_{a}^{A} (x_{a},\mu_{F}; {\tt LHAID\_A}) }{f_{b}^{B} (x_{b},\mu_{F}; {\tt LHAID\_B}) }{\hat{{\sigma}}}_{ab \to C+X}(x_{a},x_{b},\mu_{F},\mu_{R})
    \label{eq:equation2}
\end{equation}

For resolved photoproduction, we can apply Eq.~(\ref{eq:equation2}), substituting one PDF with the resolved photon PDF $f_{a}^{\gamma}$ and the other with a proton PDF. By then convoluting the result from Eq.~(\ref{eq:equation2}) with the photon flux at a bin by bin level, the photoproduction cross section for the resolved process in lepton-hadron collisions  is obtained. For an $ep$ collision, under the equivalent photon approximation~\cite{Frixione:1993yw}, we can write the resolved photopruduction cross section as: 
\begin{equation}
    \sigma_{eh\to C+X}=\sum\limits_{a,b} \int\!dx_{\gamma}dx_{a} dx_{b}{f_{\gamma}^{e} (x_{\gamma},Q^{2}_{\rm max})}{f_{a}^{\gamma}(x_{a},\mu_{F}; {\tt LHAID\_\gamma}) }{f_{b}^{h} (x_{b},\mu_{F}; {\tt LHAID\_h}) }{\hat{{\sigma}}}_{a b \to  C+X}
    \label{eq:equation3}
\end{equation} 
where $f_{\gamma}^{e}(x_{\gamma},Q^{2}_{\rm max})$ denotes the photon flux and $Q^{2}_{\rm max}$ is the maximum virtuality of photon. In the case of direct $ep$ photoproduction, one requires to replace one of the PDFs in Eq.~(\ref{eq:equation1}) with the photon flux  and compute
the cross section~\cite{Manna:2024qzf} as: 
\begin{equation}
    \sigma_{eh\rightarrow C+ X}=\sum\limits_{b} \int\!dx_{\gamma} dx_{b}{f_{\gamma}^{e} (x_{\gamma},Q^{2}_{\rm max}) }{f_{b}^{h} (x_{b},\mu_{F};{\tt LHAID\_h}) }{\hat{{\sigma}}}_{\gamma b \rightarrow C+X}(x_{\gamma},x_{b},\mu_{F},\mu_{R})
     \label{eq:equation4}
\end{equation}

\section{Validation and predictions for the EIC}\vspace*{-0.3cm}
To validate our \texttt{MG5} extension at NLO, we considered the results published in Ref.~\cite{H1:2012ffl}, in which
both resolved and direct photoproduction contribution were computed. Eq.~(\ref{eq:equation3}) is used to calculate the cross section for $b-$quark resolved photoproduction.
\begin{figure}[htbp]
\centering
  \includegraphics[width=.49\textwidth]{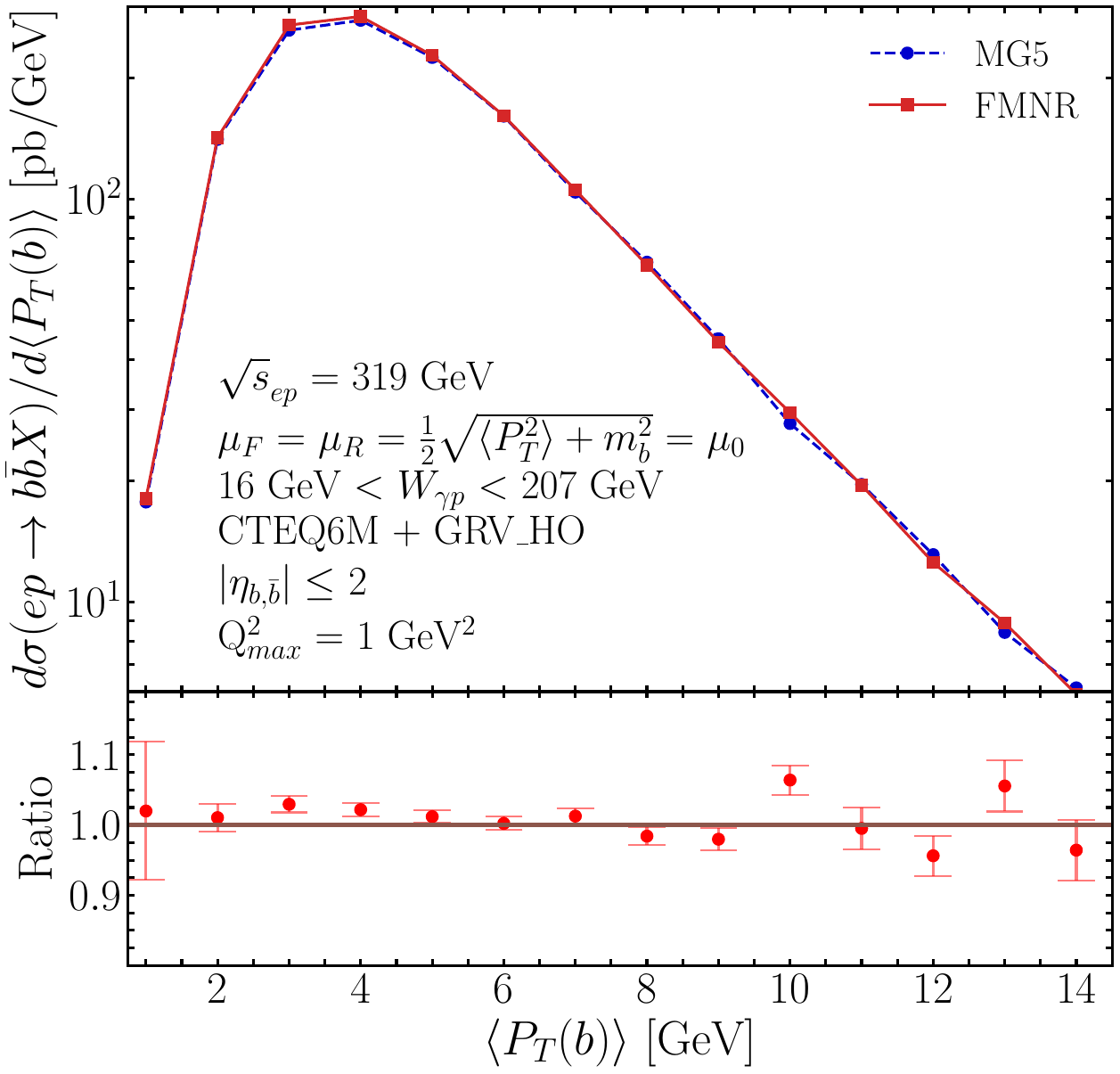}
  \includegraphics[width=.49\textwidth]{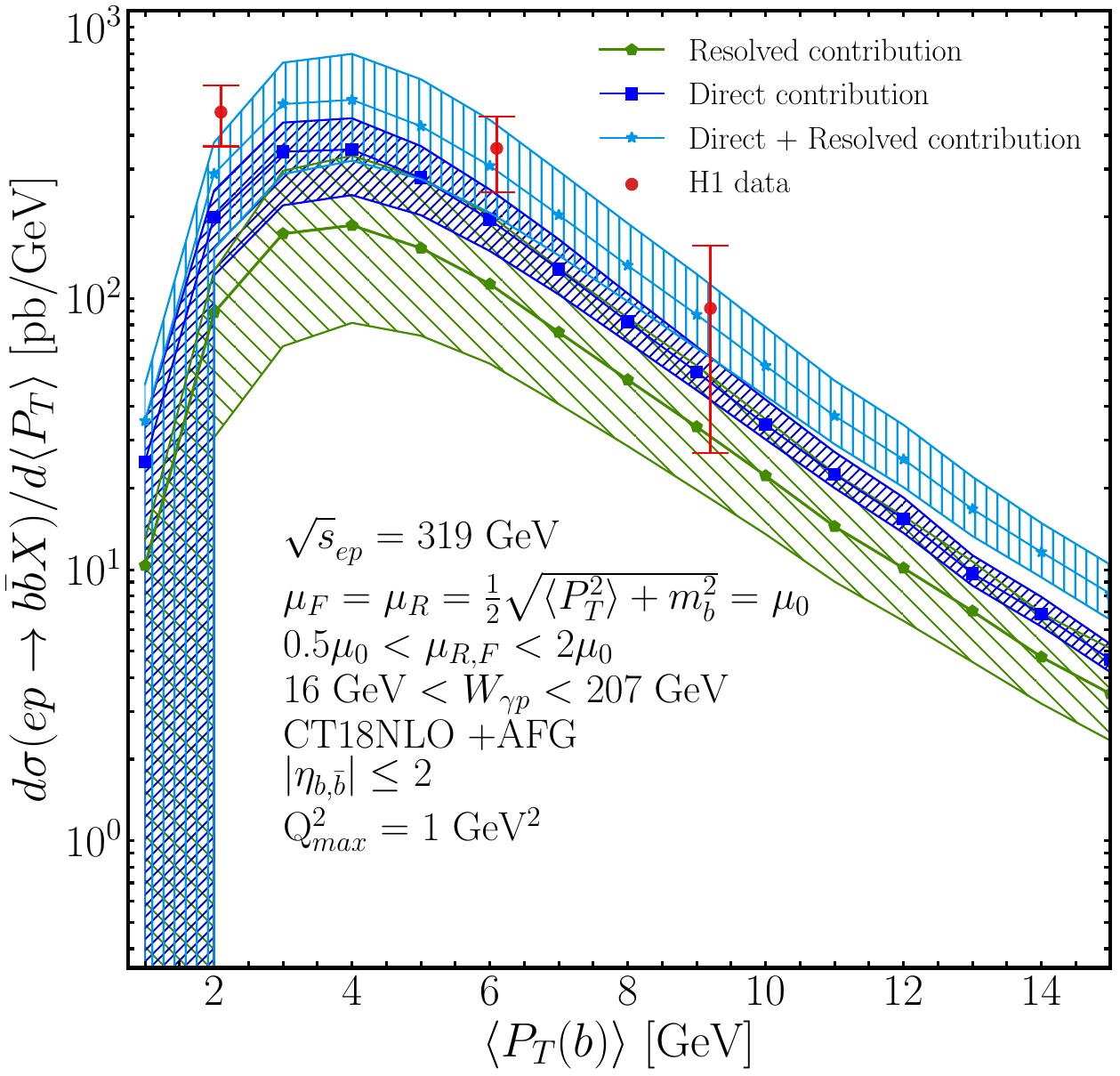}

\caption{Left: validation of {\tt MG5} for resolved photoproduction processes. The transverse momentum distribution of $b$-quark photoproduction at $\sqrt{s_{ep}}$ = 319 GeV at HERA kinematics is considered. The upper panel shows the comparison between our \texttt{MG5} computations and \texttt{FMNR}~\cite{Frixione:1993dg} predictions, while the lower panel presents the ratio of \texttt{MG5} to \texttt{FMNR}, with the combined Monte Carlo uncertainties of both codes. Right: comparison of {\tt MG5} predictions with HERA H1 data for $b$-quark production~\cite{H1:2012ffl}. The bands represents the envelope for the $\mu_F$ and $\mu_R$ uncertainties. The direct photoproduction result is obtained from Ref.~\cite{Manna:2024qzf} by using Eq.~(\ref{eq:equation4}).}
\label{fig1}
\end{figure}
Figure~\ref{fig1} (left) presents the quadratically averaged transverse momentum $\langle P_{T}(b)\rangle$ distribution, where $\langle P_{T}(b)\rangle = \sqrt{(P_{T}(b)^{2} + P_{T}(\bar{b})^{2})/2}$, for $b$-quark resolved photoproduction at a center-of-mass (CM) energy of $\sqrt{s_{ep}} = 319$ GeV. We computed the cross section for photon-proton interactions and multiplied it with the photon flux within the range 16 GeV $ < W_{\gamma p }<$ 207 GeV. We set $m_b = 4.75$ GeV, and we took $\mu_{R,F} = \mu_{0} = \frac{1}{2} \sqrt{m_{b}^{2} + \langle P_{T}(b)\rangle^{2}}$. We adopted the GRV-HO~\cite{Gluck:1991jc} photon PDF set, and the CTEQ6M~\cite{Nadolsky:2008zw} proton PDF set. According to H1 data, we applied a kinematical cut of $\lvert \eta_{b, \bar{b}}\rvert \leq 2$. Our results indicate $\sim\mathcal{O}(2\%)$ agreement between \texttt{FMNR} and \texttt{MG5} in the range $1$ GeV $< \langle P_{T}(b)\rangle < 10$ GeV. Subsequently, we compared our calculated $\langle P_{T}(b)\rangle$ spectrum for resolved photoproduction and direct one~\cite{Manna:2024qzf} with the experimental HERA H1 data presented in Ref.~\cite{H1:2012ffl}, as shown in Figure~\ref{fig1} (right panel). In summary, our results for photoproduction at HERA are in good agreement with those of \texttt{FMNR} and show consistency with experimental results of HERA H1 when both the direct and resolved contributions are considered. 

Let us now discuss cross-section computations for an upcoming $ep$ facility. Here we focus on $b$-quark production at the EIC, and present transverse momentum spectra at two different CM energies, comparing it with that of direct photoproduction~\cite{Manna:2024qzf}. 
\begin{figure}[htbp]
\centering
  \includegraphics[width=.49\textwidth]{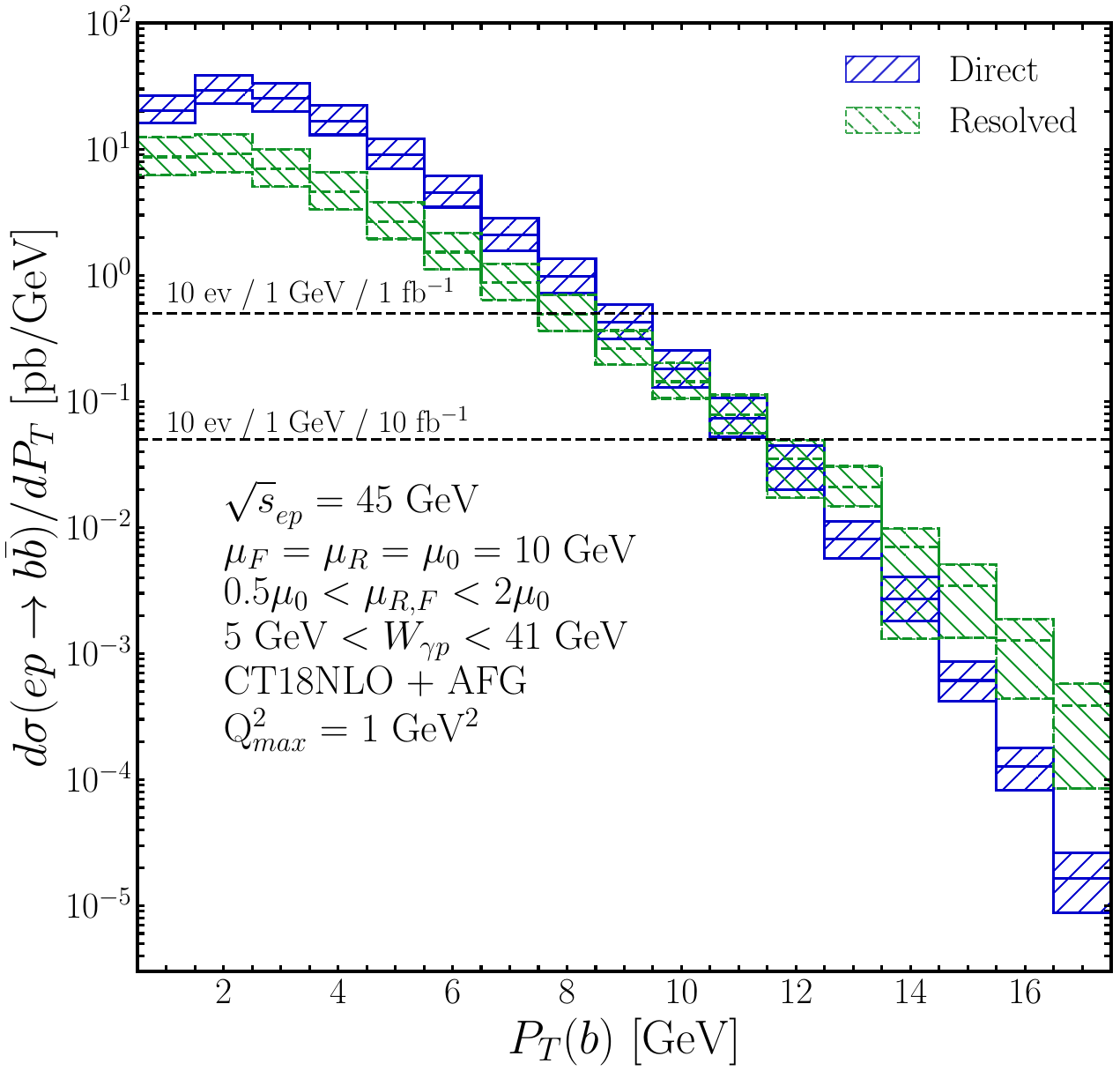}
  \includegraphics[width=.49\textwidth]{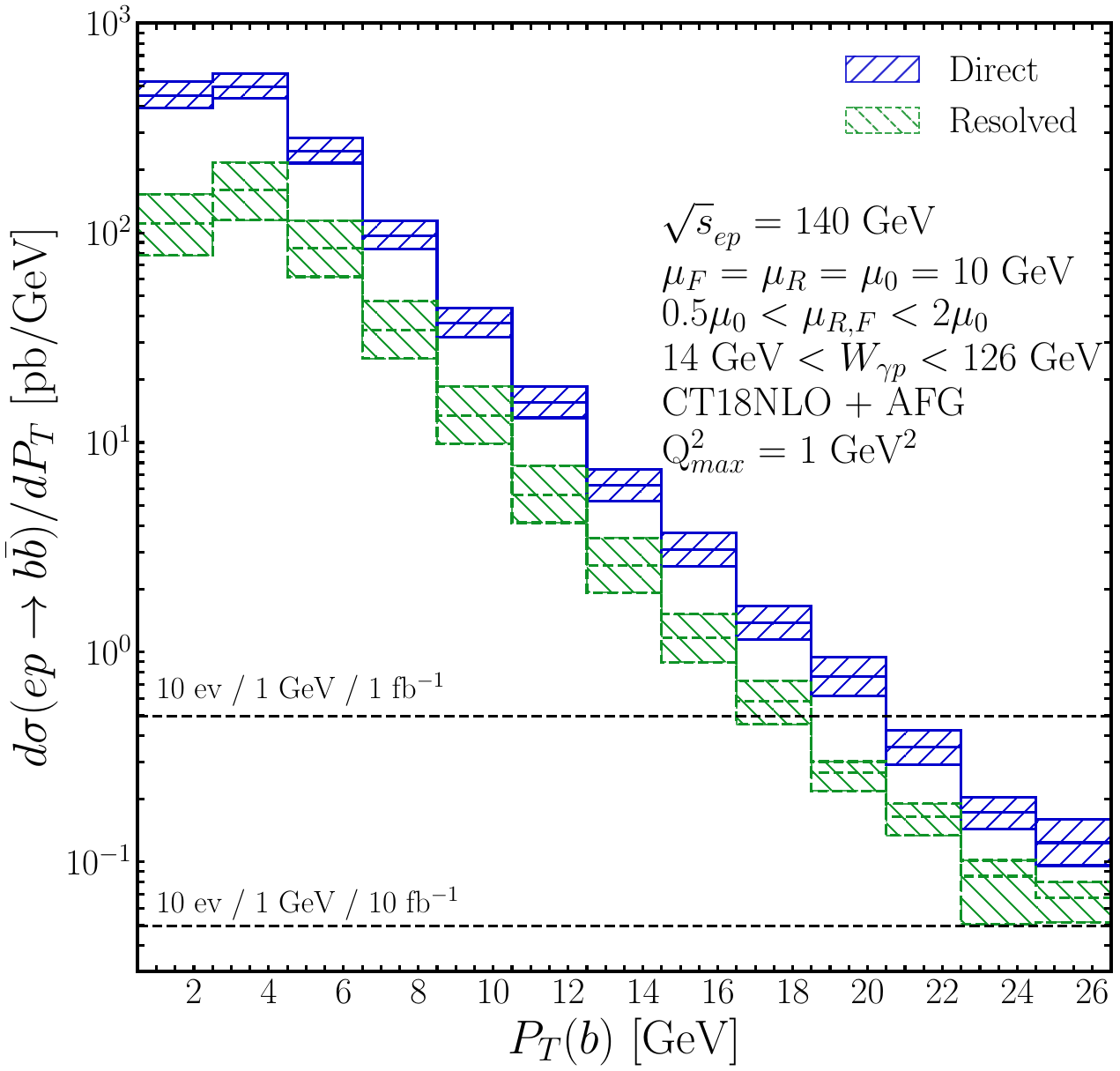}
\caption{Comparison of the transverse momentum distribution of $b$ quark direct and resolved photoproduction as predicted by \texttt{MG5} at two different energies $\sqrt{s_{ep}} = 45$ GeV  and $\sqrt{s_{ep}} = 140$ GeV. The PDF choice and the scales range are same in both cases but with different $W_{\gamma p}$ region.}
\label{fig2}
\end{figure} 
Here, we set $m_b = 4.75$ GeV and fix $\mu_R = \mu_F = 10$ GeV. We used the CT18NLO proton PDF~\cite{Hou:2019efy} and the AFG photon PDF~\cite{Aurenche:1994in}. The photon flux calculated in the range of 5 GeV $ < W_{\gamma p} < $ 41 GeV for $\sqrt{s_{ep}} $= 45~GeV and 14 GeV $ < W_{\gamma p} < $ 126 GeV for $\sqrt{s_{ep}}$ = 140 GeV. The horizontal observability lines and corresponding integrated luminosity for each energy are shown for $b \to B^0$ production, accounting for a 5\% detection efficiency which we adopted from the performance of the CMS detector at the LHC~\cite{CMS:2011pdu}. These lines indicates the $P_T$ range where $b$ quarks can be observed and highlight the regions requiring study of direct and resolved contributions. Figure~\ref{fig2} shows that, at $\sqrt{s_{ep}} = 45$ GeV, the direct photon contribution is dominant for $P_T < 12$ GeV, while resolved-photon contributions become more significant for $P_T > 12$ GeV. Conversely, at $\sqrt{s_{ep}} = 140$ GeV, direct photon processes dominate throughout the entire $P_T$ spectrum. Note that this observation may depend on the considered $W_{\gamma p}$ range.

\section{Conclusions}\vspace*{-0.3cm}
We have successfully validated the resolved-photoproduction processes within \texttt{MG5}, which now enables, among others, precise predictions for different observables, including rapidity and transverse momentum, for any hard inclusive photoproduction process across various CM energies. We anticipate that our extension of \texttt{MG5} will offer the experimental and phenomenology communities a valuable tool for exploring the potential of the EIC and other future lepton-hadron colliders, as well as for interpreting the data collected at these facilities. For example, it will help us identify the regions where the contribution from resolved photoproduction is significant.

\acknowledgments \vspace*{-0.3cm}
This work was partially supported by the Excellence Initiative: Research University at Warsaw University of Technology and the European Union’s Horizon 2020 research and innovation program under Grant Agreement No. 824093 (Strong2020), contributing to the EU Virtual Access program "NLOAccess". The project also received funding from the French Agence Nationale de la Recherche (ANR) through the grant ANR-20-CE31-0015 ("PrecisOnium") and was further supported in part by the French CNRS through the COPIN-IN2P3 bilateral agreement, and by the European Union ''Next Generation EU" program through the Italian PRIN 2022 grant n.~20225ZHA7W (C.F.). OM acknowledges support by FRS-FNRS (Belgian National Scientific Research Fund) IISN projects 4.4503.16.

\bibliographystyle{JHEP}\small
\bibliography{DIS2024/reference}

\end{document}